\documentclass[conference]{IEEEtran}
\IEEEoverridecommandlockouts

\usepackage{cite}
\usepackage{url}
\usepackage{amsmath,amssymb,amsfonts}
\usepackage{algorithmic}
\usepackage{graphicx}
\usepackage{textcomp}
\usepackage{xcolor}
\usepackage{makecell} 
\def\BibTeX{{\rm B\kern-.05em{\sc i\kern-.025em b}\kern-.08em
    T\kern-.1667em\lower.7ex\hbox{E}\kern-.125emX}}
\begin{document}

\title{Auditable Session Admission for Cross-Silo Federated Learning
 
\thanks{Accepted at the 2nd International Conference on Federated Learning and
Intelligent Computing Systems (FLICS2026)}
}

\newif\ifANONYMOUS  
\ANONYMOUSfalse 
 
\ifANONYMOUS
\else
  \author{\IEEEauthorblockN{Enzo Fenoglio}
  \IEEEauthorblockA{\textit{dept.Computer Science} \\
  \textit{University College London}\\
  66-72 Gower Street, London WC1E 5EA, UK\\
  e.fenoglio@ucl.ac.uk}
  \and
  \IEEEauthorblockN{Philip Treleaven}
  \IEEEauthorblockA{\textit{dept.Computer Science} \\
  \textit{University College London}\\
  66-72 Gower Street, London WC1E 5EA, UK\\
  p.treleaven@ucl.ac.uk}
  }
\fi

\maketitle

\begin{abstract}
Cross-silo federated learning keeps raw data local, but deployments frequently stall on a practical bottleneck when deciding who may invoke which session-scoped operations across organizational boundaries, under constraints that remain auditable after execution. In practice, admission is implemented via centralized policy services, platform configuration, or ad hoc checks, which drift over time and are hard to audit from boundary-visible evidence. We present a session admission gateway that enforces pre-approved session capabilities at the boundary and emits verifiable decision records. During session setup, participating organizations approve roles and constraints and mint signed session capability tokens that enumerate permitted session operations for a given \textit{session\_id}. At runtime, each request carries the token and a request-bound proof-of-possession, making replay and impersonation with stolen tokens detectable at the gateway via request binding. Admission reduces to stateless per-request cryptographic verification and capability matching at an admission gateway, while setup and orchestration remain out of scope and organization-specific. We validate the approach end-to-end on a cross-silo FL workflow using MNIST as a surrogate workload. The proof-of-concept is open-source, containerized, provisioned as code, and includes reproducible tests and evidence logs.
\end{abstract}

\begin{IEEEkeywords}
Session admission; JOSE; proof-of-possession; cross-silo federated learning; auditability.
\end{IEEEkeywords}

\section{Introduction}
Federated learning (FL) enables multiple organizations to train a shared model without centralizing raw data. In cross-silo settings—e.g., hospitals, banks, industrial consortia—FL is often presented as a practical path to collaboration under data-locality and regulatory constraints. However, deployments can stall or become brittle when participants must decide who is allowed to invoke which session-scoped operations across organizational boundaries—particularly in cross-silo FL, where multiple organizations must pre-approve and later justify join, fetch, submit, and evaluation actions under a shared training session. These governance and administrative questions are repeatedly identified as adoption barriers and deployment-critical concerns in cross-organization FL~\cite{Barbereau_2025, Sprenkamp_2024 }, especially in regulated domains such as healthcare~\cite{Kommusaar_2026, xu_2021}.  

The gap between adoption and deployment is most visible at the FL session interface for creating and running a training session, joining participants, fetching the current model, submitting updates, and evaluating results. In practice, session admission is frequently implemented through a patchwork of cloud Identity Access Management (IAM) policies, platform-specific configuration, centralized policy services, or ad hoc checks embedded in orchestration code. These are useful building blocks for authentication, infrastructure access, and internal authorization, but they do not provide portable, session-scoped admission semantics with boundary-verifiable evidence. In cross-silo deployments, they degrade for three reasons. First, they are non-portable because each organization’s identity, policies, and deployment stack differ, making integration bespoke. Second, they are hard to audit as decisions are distributed across configurations and services, making it difficult to reconstruct why an operation was allowed for a given session. Third, they are brittle under drift as operational changes can silently alter admission behavior without a durable decision record tied to the run. Our contribution targets session admission and audit evidence at the session interface boundary; privacy and robustness mechanisms for training and aggregation are orthogonal and can be layered on top.    

We distinguish cross-silo session admission from in-session authorization. The difference is similar to that between a passport and a badge: the passport determines whether a participant may enter the session, whereas the badge corresponds to organization-local IAM governing what that participant may do once inside. Session admission governs the cross-boundary transition from session setup (enrollment, approvals, quorum formation) to session execution, where multiple organizations must agree on and enforce session-scoped permissions for session interface operations. Once admitted, each organization may still apply its conventional, stateful IAM (RBAC/ABAC and internal service authorization) to its own users, services, and data during session execution. Federated identity and cross-account IAM can authenticate principals and grant infrastructure-level access, but they do not define session-scoped, operation-level admission semantics for FL together with execution-time evidence that remains verifiable after the fact. Our goal is not to replace local IAM within each organization, but to supply an admission substrate at the cross-silo boundary.

This paper proposes a session admission mechanism for cross-silo FL by making each session-interface request proof-carrying. Each request includes (i) a signed \textit{session capability token (SCT)} encoding the permitted session operations for a specific \textit{session\_id}, and (ii) a request-bound holder-key proof. Admission is enforced at the session boundary by an admission gateway on the request path. Unlike approaches where authorization is embedded in platform configuration or orchestration logic (Section~\ref{sec:related}), the proposed design enforces admission as stateless per-request verification at the session boundary, using request-carried evidence and producing explicit decision records.

Organizations approve participation and administrators mint SCTs; the roles encoded in the SCT determine the permitted operations (e.g., administrative roles for lifecycle actions such as \textit{start\_session} and \textit{end\_session}, and participant roles for session operations such as \textit{fetch\_model}, \textit{submit\_update}, and \textit{evaluate}). Each admission decision emits a structured record (operation, \textit{session\_id}, reason code, timestamp, and hashes of the token and request) that is appended to an evidence log and correlated with run artifacts, enabling later audit without reconstructing transient platform configuration. The contribution is an FL-facing integration of standard components—multi-party authorization, signed capability tokens, and request-bound key proofs—into an admission gateway with auditable decision records.

We demonstrate the approach in a containerized cross-silo FL prototype, augmented with an admission gateway and evidence logging. The gateway enforces session-scoped permissions at the boundary and records structured evidence for both admitted and denied requests, including negative tests for capability-mismatch and possession-mismatch cases. Model artifacts are persisted in an artifact store and are accessible only under the appropriate session capabilities. The prototype reuses infrastructure from a broader federated-computing testbed~\cite{fenoglio_fcac_2026, fenoglio_rsos_2026}; here we isolate and evaluate only the FL-facing session admission layer and its operational behavior. In summary, we make the following contributions:
 
\begin{enumerate}
    \item \textbf{Session-boundary admission for cross-silo FL.} We specify an FL-native admission layer in which each session-interface request is authorized via signed session capabilities bound to a \textit{session\_id} and verified at the boundary.
    
    \item \textbf{Proof-carrying requests.} We bind each session-interface request to a holder key via proof-of-possession, making stolen tokens insufficient on their own and enabling per-request admission at the session boundary.
    
    \item \textbf{Auditable decision records and reproducible evidence.} We emit structured decision records and correlate them with run artifacts in a reproducible Flower-based prototype, including negative tests for capability and possession mismatch.
\end{enumerate}

The remainder of the paper is organized as follows. Section 2 discusses related work. Section 3 presents the system and threat model. Section 4 specifies the session admission mechanism and verification procedure. Section 5 describes the separation of the session lifecycle into setup/orchestration and runtime admission. Section 6 reports the prototype and experimental evidence. Section 7 discusses limitations and operational considerations, and Section 8 concludes.

\section{Related Work}
\label{sec:related}

Cross-silo FL deployments are often constrained by governance and authorization at the session boundary. Who may create/join a session, which participants may invoke which session operations, under what constraints, and with what evidence those decisions can be justified after execution. Empirical and survey-style accounts repeatedly identify governance, administrative coordination, and auditability as deployment-critical barriers in cross-organization FL. This motivates enforcing session-interface permissions through explicit boundary enforcement points and deterministic checks over request-carried evidence. This boundary-centric stance aligns with Zero Trust Architecture as standardized by NIST, which shifts security away from implicit perimeter trust toward explicit, policy-driven access decisions applied at well-defined enforcement points~\cite{Rose_2020}. NIST guidance operationalizes this model for cloud-native deployments, emphasizing application-level access control and runtime enforcement~\cite{Chandramouli_2023}. Our session admission layer is consistent with this pattern when applied to cross-silo FL session interfaces. Commercial federated-computing platforms provide practical controls such as RBAC and audit logs, but admission semantics often remain platform-local~\cite{rhino_fcp_platform_2026}; we instead externalize session-interface admission as request-carried evidence checked at the boundary.

\subsection{Complementary identity, IAM, and zero-trust mechanisms}
Federated identity management (e.g., SAML and OIDC) and authorization frameworks such as OAuth are widely used building blocks for cross-organization deployments. They establish authentication, propagate identity attributes, and support delegated authorization for APIs and services~\cite{saml2core,oidc_core,oauth2_rfc}. Similarly, cross-account or cross-tenant IAM in cloud platforms provides practical \emph{infrastructure-level} authorization across organizational boundaries (e.g., granting access to storage, compute, or service principals). These mechanisms are often necessary foundations for deployment, but they do not define a \emph{multiparty session admission contract} for \emph{session-scoped, operation-level} control of cross-silo FL session interfaces, nor do they yield boundary-verifiable decision evidence tied to a specific run. In practice, session admission logic is pushed into provider-specific configuration and orchestration code, where it becomes non-portable, and difficult to justify from boundary-visible evidence post hoc~\cite{Barbereau_2025}. Our admission layer is designed to compose with these foundations by externalizing session admission semantics as signed capability artifacts verified at the session boundary and recorded as structured decision evidence (Section~\ref{sec:mechanism}).  

Zero-trust architectures provide applicable design principles — continuous verification and least privilege — but they are largely agnostic to the application-level contract of a given system. In FL, the gap is twofold: (1) an explicit mapping from session requests to session-scoped permissions and constraints, and (2) boundary-verifiable evidence of each access decision tied to a specific run. We address both gaps with a session admission gateway that enforces session-scoped authorization at the session boundary while leaving identity and key management organization-local.

\subsection{Capability-based authorization and proof-of-possession}
Capability-based access control and signed authorization artifacts have a long history, including SPKI/SDSI-style credentials for delegated authorization~\cite{rfc2693}. More recent capability systems such as Macaroons demonstrate practical attenuation and delegation patterns suitable for distributed services~\cite{macaroons2014}. Proof-of-possession mechanisms address the shortcomings of bearer tokens by binding tokens to a key held by the requester; in the OAuth ecosystem this includes DPoP~\cite{rfc9449} and related token-binding approaches. In our setting, these established mechanisms are combined into a stateless admission layer for the FL session interface.

\section{System and Threat Model}
\label{sec:system}

\subsection{Setting and roles}
We consider cross-silo FL among multiple organizations (e.g., Org A and Org B) that collaborate to train a shared model while keeping raw data local. Each organization operates FL clients that train on local data and interact with an FL server (or server-side service) that coordinates rounds and maintains the global model. 

\paragraph{Administrative roles}
\emph{Organization administrators} authorize participation and mint session-scoped authorization artifacts for their organization. \emph{Regular users} (or client processes acting on their behalf) participate in a session via a restricted set of session operations such as \textit{fetch\_model}, \textit{submit\_update}, and \textit{evaluate}. Session lifecycle operations (e.g., \textit{start\_session}, \textit{end\_session}) are administrator-controlled.

\paragraph{Member identity and registration}
Within each organization, participants are identified by an organization-scoped member identity (\texttt{sub}) and an optional holder-key binding (\texttt{cnf.jkt}) used for PoP verification. The prototype realization of member registration is described in Section~\ref{sec:prototype}-B.

\subsection{Session interface}
\begin{figure}[t]
\centering
\includegraphics[width=\columnwidth]{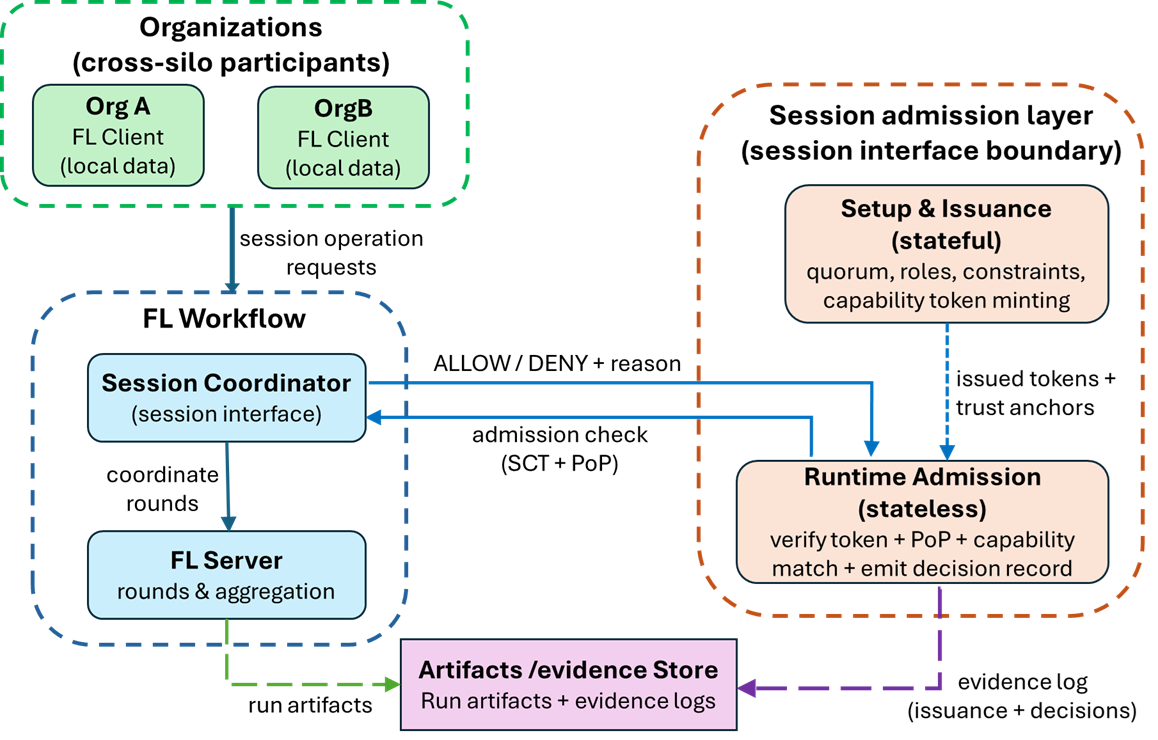}
\caption{Cross-silo FL workflow with a session admission layer at the session interface boundary. Setup/issuance produces organization-issued Session Capability Tokens SCTs; runtime requests are checked at the admission gateway and yield \texttt{ALLOW}/\texttt{DENY} decisions with logged evidence.}
\label{fig:architecture}
\end{figure}

We use the term \emph{session interface} to denote the set of session-scoped endpoints/RPCs that create and operate an FL training session across participants. We include session lifecycle actions (e.g., start/join/end) and participant operations such as \textit{fetch\_model}, \textit{submit\_update}, and \textit{evaluate}. Our admission layer fronts this interface across organizations; the underlying training and aggregation procedures are unchanged. Figure~\ref{fig:architecture} provides an overview of the workflow and the placement of the session admission layer that is zoomed in Figure~\ref{fig:admission_zoom}; Section~\ref{sec:mechanism} details the protocol.

\subsection{Admission artifacts}
Our mechanism uses three request-relevant elements: (i) a session identifier \texttt{session\_id} binding authorization and evidence to a specific run; (ii) an SCT enumerating the session operations permitted for its holder for that \texttt{session\_id}; and (iii) a request-bound proof-of-possession signature demonstrating holder-key control over the request. Section~\ref{sec:mechanism}-B specifies how SCT claims (e.g., \texttt{sub} and \texttt{cnf.jkt}) bind the token to the holder key used for PoP verification~\cite{rfc8705, rfc7800}. Organization-specific policies and participation constraints are evaluated during session setup and compiled into the claims of the issued SCT. As a result, runtime admission depends only on the token contents and request-bound proof, without reliance on external policy state.

\subsection{Evidence and artifact storage}
Each admission decision emits a structured JSON decision record (e.g., token hash, canonical request hash, \texttt{session\_id}, decision, reason, timestamps). Run artifacts (metrics and model checkpoints) are persisted to an artifact store. Model artifacts are not assumed to be publicly readable; access is mediated by session capabilities, so that possessing run evidence does not imply the right to retrieve model checkpoints.

\subsection{Threat model and non-goals}
\label{sec:threat_model}
\paragraph{Operational setting}
We target cross-silo deployments in honest-but-fallible environments where heterogeneous identity/IAM stacks, configuration drift, and integration dependencies are common sources of authorization errors and audit gaps.

\paragraph{Adversary and attacks}
An adversary may attempt to (i) invoke session operations without authorization, (ii) replay a stolen SCT, (iii) impersonate another principal or organization, or (iv) misuse a valid SCT by invoking an operation outside its permitted capability set or outside its bound \texttt{session\_id}. Our mechanism makes each session-interface request verifiable at the boundary via token validation, request-bound PoP, and capability matching. Table~\ref{tab:evidence}  summarizes how each check addresses a specific attack vector and the evidence fields it produces.

\paragraph{Cryptographic assumptions}
We assume standard signature primitives are secure and that organization-managed private keys are protected by a vault; key compromise, revocation, and recovery are treated as limitations (Section~\ref{sec:limitations}).

\paragraph{Non-goals}
We do not address privacy or robustness mechanisms for the FL learning procedure itself (e.g., secure aggregation, differential privacy, or defenses against malicious model updates). These mechanisms are complementary and operate within the learning procedure after requests have been admitted, whereas our focus is the deployability and auditability of cross-silo authorization for the session interface.

\paragraph{Failure semantics}
A \texttt{DENY} decision blocks only the requested operation; it does not terminate the session. A session may be created once quorum is reached and remain inactive until permitted operations occur (Section~\ref{sec:lifecycle}-B).

\begin{figure*}[t]
\centering
\includegraphics[width=0.55\linewidth]{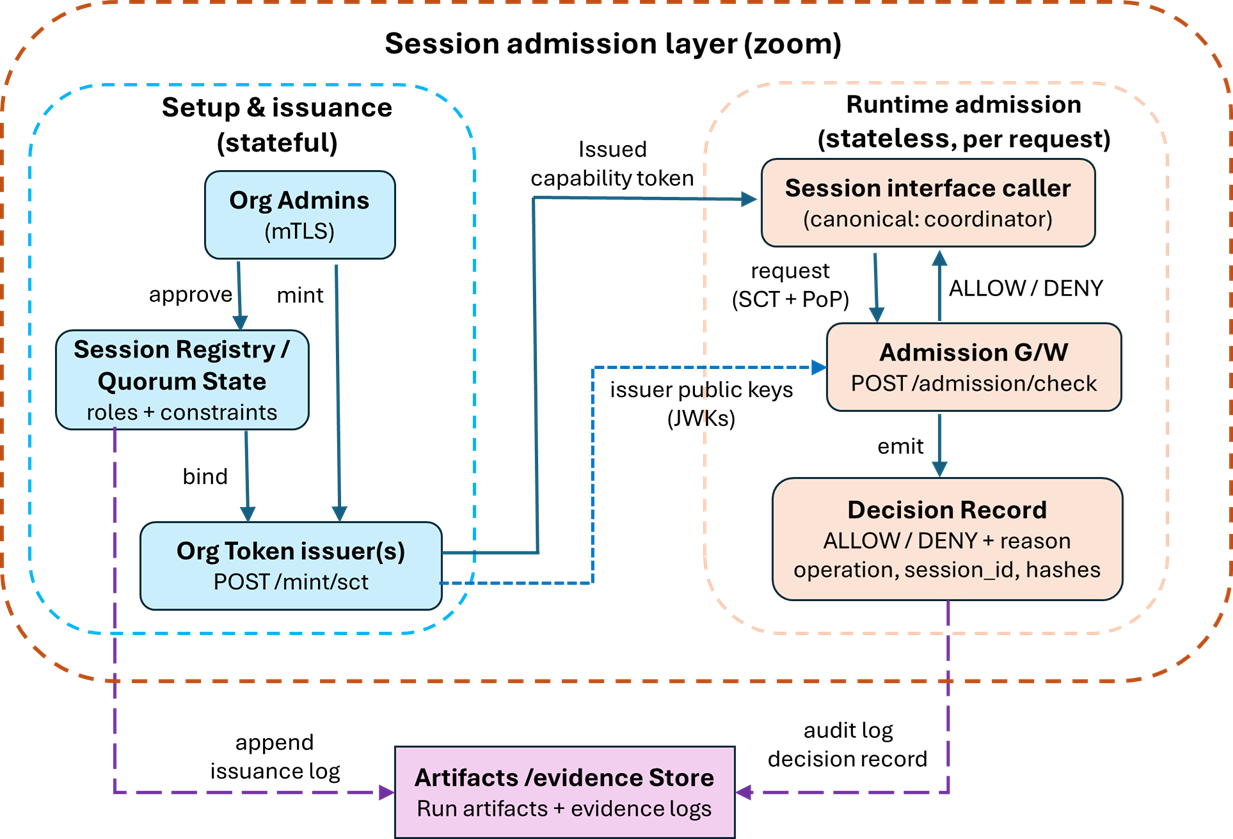}
\caption{Session admission layer (Zoom). During setup, administrators approve quorum roles/constraints and mint SCTs. At runtime, the coordinator submits session requests carrying the SCT  and PoP; the admission gateway returns \texttt{ALLOW}/\texttt{DENY}, emits a decision record (\textit{operation}, \texttt{session\_id}, hashes), and appends evidence to the artifact/evidence store.}
\label{fig:admission_zoom}
\end{figure*}

\section{Session Admission Mechanism}
\label{sec:mechanism}

\subsection{Design goals}
We aim to make cross-silo authorization for the \emph{session interface} (i) \emph{portable} across heterogeneous organizational stacks, (ii) \emph{verifiable at the boundary} from request-carried evidence, and (iii) \emph{auditable after execution} via run-tied decision records.
The mechanism is designed to compose with federated identity and cloud IAM rather than replace them. Identity and infrastructure access may remain organization and provider-specific, while session-scoped authorization is enforced uniformly at the admission gateway. Figure~\ref{fig:admission_zoom} illustrates the split between stateful setup and per-request runtime admission. Conceptually, the chain separates trust-anchor configuration, issuance of session-scoped permissions, and per-request holder-key binding, enabling deterministic cryptographic verification at the session boundary.

\subsection{Session capability tokens}
A \emph{session capability token (SCT)} is a signed artifact minted during session setup by an organization administrator. We assume administrators authenticate using mutual TLS (mTLS) over TLS~\cite{rfc8446}; OAuth mutual-TLS profiles provide one concrete realization when OAuth is used~\cite{rfc8705}. An SCT binds permissions to a specific session and to a holder identity under the issuing organization trust domain. Concretely, the token contains at least:
\begin{itemize}
  \item \texttt{session\_id}: session identifier;
  \item \texttt{issuer}: organization identifier (e.g., Org A);
  \item \texttt{sub}: holder identifier within the issuing organization domain;
  \item \texttt{cnf.jkt}: holder-key binding (thumbprint of the holder public key) used for PoP verification;
  \item \texttt{capabilities}: a set of permitted session operations (e.g., \{\textit{fetch\_model}, \textit{submit\_update}, \textit{evaluate}\});
  \item \texttt{constraints}: optional predicates (time window, round bounds, rate limits, etc.);
  \item \texttt{exp}: expiration/validity interval.
\end{itemize}
We use \texttt{sub} to identify the holder within the issuing organization’s namespace, and \texttt{cnf.jkt}~\cite{rfc7800, rfc7638} to bind the token to the holder’s PoP verification key by carrying the thumbprint of the holder public key. The SCT is signed by an organization-managed key (stored in a secure vault), and the admission gateway is configured with the corresponding trust anchors/public keys for participating organizations.

\paragraph{Realization with standard token formats}
SCTs can be realized using the JOSE (JSON Object Signing and Encryption) framework: a signed JSON Web Token (JWT) carrying the session claims and capability set, signed using the  JSON Web Signature (JWS), with issuer public keys distributed as JSON Web Keys (JWKs)~\cite{rfc7519,rfc7515,rfc7517}. If deployments require explicit holder-of-key semantics, standardized confirmation claims can associate the token with a proof key~\cite{rfc7800}. When stable key identifiers are needed for verification material selection, JWK thumbprints provide a standard, collision-resistant identifier~\cite{rfc7638}. JWT best-current-practice guidance applies to avoid common validation pitfalls~\cite{rfc8725}. We remain format-agnostic at the abstraction level; these references illustrate a widely deployed realization.

\subsection{Request-bound proof-of-possession}
Bearer tokens alone are vulnerable to replay and impersonation when stolen. Each request includes a \textit{proof-of-possession} (PoP) signature under the holder’s private key over a canonical representation of the request, so a stolen SCT alone cannot be used to invoke session operations. We adopt the DPoP approach from RFC~9449~\cite{rfc9449}, signing a JWT that includes:
\begin{itemize}
  \item HTTP method and endpoint (or RPC method name),
  \item \texttt{session\_id},
  \item a nonce or unique request identifier (\texttt{jti}),
  \item a timestamp (and optional TTL),
  \item a hash of the request body where applicable.
\end{itemize}
The admission gateway verifies that the PoP signature corresponds to the holder identity asserted by the SCT (via \texttt{cnf.jkt}) and that the signed claims bind to the request being evaluated. This prevents token replay even if an SCT is intercepted, as the attacker cannot forge valid request-specific signatures without the holder's private key.

\paragraph{Canonical hashing}
To ensure that request hashes and evidence fields are stable across implementations, JSON inputs are canonicalized before hashing. We apply the JSON Canonicalization Scheme (JCS, RFC~8785~\cite{rfc8785}) to request bodies and decision records, providing deterministic request/evidence hashing regardless of key ordering, whitespace, or number formatting.

\subsection{Admission gateway procedure}
The admission gateway fronts the session interface and returns \texttt{ALLOW} or \texttt{DENY} for each request. In our architecture, enforcement is canonical at the coordinator boundary (Figure~\ref{fig:admission_zoom}). The coordinator invokes the admission check for session-scoped operations it forwards to the session interface; clients may optionally perform local sanity checks (e.g., token expiry and operation presence) for usability without being authoritative.  Verification consults only request-carried evidence and configured trust anchors; a bounded replay cache (optional) is an implementation detail. Verification consists of:
\begin{enumerate}
  \item \textbf{SCT validation:} verify signature, issuer trust anchor, expiration, and required fields; parse \texttt{session\_id}, capability set, and any constraints.
  
  \item \textbf{Capability match:} map the requested endpoint/RPC to an abstract session operation and check that it is permitted by the SCT for the bound \texttt{session\_id}; evaluate any token constraints.
  \item \textbf{PoP validation:} reconstruct the canonical request string and verify the PoP signature; check nonce/timestamp freshness to reduce replay.
  \item \textbf{Decision record:} emit a structured record (Section~\ref{sec:system}) with \texttt{ALLOW}/\texttt{DENY} and a reason code.
\end{enumerate}
A denied request blocks only the requested operation. The session may continue for other participants and operations that satisfy their capabilities.

\begin{table*}[t]
\caption{Session admission checks, addressed failure modes, and minimal evidence fields for audit justification.}
\centering
\small
\begin{tabular}{|c|c|c|}
\hline
\textbf{Admission check} & \textbf{Prevents/detects} & \textbf{Evidence fields} \\
\hline
Token signature \& validity & Forged/expired tokens & token hash, \texttt{iss}, \texttt{exp}, signature verification outcome, decision \\
PoP verify & Impersonation with stolen token & request hash, PoP verification outcome \\
PoP freshness (nonce/time) & Replay of previously valid requests & \texttt{jti}, timestamp, freshness outcome \\
Capability match (operation) & Unauthorized session operation & operation, capability set id/hash, reason code \\
Session binding & Cross-session token misuse & \texttt{session\_id}, token \texttt{session\_id} claim \\
\hline
\end{tabular}
\label{tab:evidence}
\end{table*}

\section{Session Lifecycle: Setup vs. Runtime Admission}
\label{sec:lifecycle}

\subsection{Separation of stateful setup from per-request admission}
A core design choice is the separation between:
\begin{itemize}
  \item \textbf{Session setup and orchestration (stateful):} enrollment workflows, approval/quorum formation, and minting of organization-issued SCTs;
  \item \textbf{Runtime admission (stateless per request):} boundary checks at the admission gateway using request-carried evidence (SCT + PoP) and configured trust anchors.
\end{itemize}
This split confines mutable coordination state (quorum, roles, constraints) to setup, but keeps runtime authorization at the session boundary (Section~\ref{sec:mechanism}).

\subsection{Quorum and session creation}
We assume a session is created once a required quorum of organizations has been reached via an administrative workflow.
Session creation establishes \texttt{session\_id} and the approved roles/constraints; administrators then mint SCTs bound to that identifier for the permitted operations. A newly created session may be \emph{inactive}, i.e., no training rounds execute until permitted participants invoke session operations consistent with their minted capabilities. Once quorum is reached and setup artifacts are available, the orchestrator initiates training. The admission gateway regulates access to session-scoped operations throughout execution by evaluating request-carried evidence.

\subsection{Role-scoped operations}
We separate administrative lifecycle operations from participant operations to reduce privilege scope and simplify review. In the prototype described in Section~\ref{sec:prototype}, administrators mint SCTs and manage lifecycle actions. Regular users are limited to participant operations (e.g., \textit{fetch\_model}, \textit{submit\_update}, \textit{evaluate}). Server-side aggregation is not exposed as a regular-user operation.

\subsection{Operational mapping to session interface endpoints}
Session interface endpoints (or RPC methods) are mapped to abstract operations used for capability matching at the gateway. This mapping is explicit and versioned with the gateway configuration to avoid implicit coupling to orchestration code. For example, \texttt{/session/model} maps to \textit{fetch\_model}, \texttt{/session/update} maps to \textit{submit\_update}, and \texttt{/session/evaluate} maps to \textit{evaluate}. Administrative endpoints map to lifecycle operations (e.g., \textit{start\_session, end\_session}). The SCT capability set is evaluated against these abstract operations rather than raw URLs to keep the policy stable under API refactors.

\subsection{Decision evidence as a run artifact}
Each admission decision appends a structured decision record to the evidence log, including \texttt{session\_id}, operation, decision (\texttt{ALLOW}/\texttt{DENY}), a reason code, and hashes of the SCT and the request (under the canonicalization used for PoP). This log can be correlated with run artifacts (model checkpoints and metrics) for later inspection without reconstructing transient IAM configuration or orchestration state.

\section{Prototype and Evidence}
\label{sec:prototype}

\subsection{Implementation overview}
We evaluate the session admission layer in a containerized cross-silo FL deployment built on a standard FL framework and a canonical image-classification workload. The deployment includes (i) an FL server and per-organization FL clients, (ii) a coordinator/orchestrator that manages session progression, (iii) an admission gateway fronting the session interface and performing stateless verification (Section~\ref{sec:mechanism}), and (iv) evidence and artifact stores for decision records and run outputs.

The proof-of-concept is packaged as containers and provisioned as code with OpenTofu~\cite{opentofu_docs} to support reproducible deployment across organizations. The FL framework is Flower~\cite{beutel2020flower}, and the workload is MNIST~\cite{lecun1998mnist}, used as a controlled surrogate to exercise session operations, admission decisions, and evidence generation end-to-end. We model organizational cohorts by partitioning digits across participants (e.g., even/odd or subsets), treating \texttt{digit} as a stand-in for a governed data category (analogous to a clinical cohort label) so that capability rules can permit or deny specific evaluation/prediction requests. We release a reproducible proof-of-concept implementation and test on GitHub\footnote{\url{https://github.com/onzelf/FLICS-cross-silo-admission}}. Evaluation on domain workloads and heterogeneous cross-organizational datasets is left to future works. 

Organization keys for SCT signing and member PoP are managed in organization-controlled vaults. Our prototype simulates vault storage on disk; production deployments would use hardware-backed key management services (e.g., AWS Nitro Enclaves, Azure Key Vault, or HSM-backed vaults). The admission gateway is configured with the corresponding public keys and trust anchors for participating organizations. This matches common operational practice in cross-silo deployments, where each party manages its own signing material and shares only the verification material required for cross-organization admission.

\subsection{Experimental setup and session interface}
Our objective is to validate that session admission is (i) enforceable at the session boundary, (ii) portable across organizations without shared runtime policy services, and (iii) auditable after execution from boundary-visible evidence. We therefore focus on session-scoped operations exposed by the session interface, rather than on learning performance.

We instantiate the two roles from Section~\ref{sec:system}-A in the prototype. Administrators control session lifecycle operations and mint SCTs during setup. Regular users are restricted to participant operations (\textit{fetch\_model}, \textit{submit\_update}, \textit{evaluate}); server-side aggregation is not exposed as a regular-user operation.

\paragraph{Member registration (prototype prerequisite)}
Before SCT issuance, each organization registers members by recording \{\texttt{org\_id}, \texttt{member\_id}, public key, \texttt{jkt}\} in an organization-local registry, where \texttt{member\_id} serves as the member's identity within that organization's namespace and \texttt{jkt} is the thumbprint of the member's public key. SCTs reference the registered member via \texttt{sub} (which equals \texttt{member\_id} in the issuing organization's registry) and carry the corresponding holder-key binding as \texttt{cnf.jkt}; PoP proofs are generated using the corresponding private key held by the member (e.g., in an organization-managed vault). This avoids relying on free-form usernames in the UI while preserving local key custody.

\subsection{Test protocol}
We test whether the admission gateway enforces session-scoped permissions and binds each request to the holder key via PoP. The protocol is structured as a small suite of end-to-end tests that exercise both success and failure modes commonly observed in cross-silo deployments.

\textit{Session setup and issuance:} Administrators approve participation constraints and mint SCTs for a shared \texttt{session\_id} (Section~\ref{sec:system}-C). We issue participant-role and administrative-role tokens to exercise participant operations versus lifecycle operations in the tests. All SCTs are signed with organization-managed keys.

\begin{enumerate}
\item \textbf{Positive path.}
Clients invoke \textit{fetch\_model} and \textit{submit\_update} with a valid SCT and a request-bound PoP. The gateway returns \texttt{ALLOW} and forwards the request to the session interface. Training proceeds and produces standard run metrics and model artifacts.

\item \textbf{Negative case 1 (capability mismatch).}
A client attempts an operation that is not included in its capability set, such as calling \textit{evaluate} using an SCT that omits \textit{evaluate}. The gateway returns \texttt{DENY} with a capability-mismatch reason code.

\item \textbf{Negative case 2 (PoP mismatch / replay).}
A request reuses an SCT with an invalid or stale PoP (e.g., wrong holder key, incorrect binding to the canonical request, or failed freshness check). The gateway returns \texttt{DENY} with a mismatch or freshness failure reason, aligning with standardized mechanisms designed to detect replay of stolen bearer material~\cite{rfc9449}.
\end{enumerate}

These tests deliberately mirror operational failure modes that are difficult to diagnose when authorization is embedded in configuration or orchestration logic. The goal is not cryptographic novelty, but to demonstrate that a session admission layer makes such failures explicit, localizable, and auditable at the session boundary.

\subsection{Findings and evidence}
The primary result is that session admission can be enforced at the session boundary and reconstructed after execution from decision records. The evaluation focuses on validating the correctness and auditability of stateless admission at the session boundary, rather than on domain realism or large-scale performance. Requests are admitted only when SCT validation, PoP verification, and capability matching succeed. The gateway denies requests that fail these checks before they reach the session interface, returning reason codes for capability mismatches or PoP failures. For each request, the evidence log records a JSON decision including \texttt{session\_id}, operation name, canonical request hash, SCT hash (optionally with issuer/subject metadata), and outcomes for PoP verification and freshness checks. Each record also includes the decision, reason code, timestamps, and a correlation identifier linking to run artifacts. Run artifacts comprise a \texttt{run.json} record (session and workload configuration), training metrics, and model checkpoints. Checkpoint retrieval is treated as a capability-checked session operation. Table~\ref{tab:evidence} summarizes the checks, failure modes, and minimal evidence fields used for post hoc admission justification.

\subsection{Overhead and deployability considerations}
The admission gateway performs per-request cryptographic verification and capability matching, which is independent of the number of session participants (per-request verification and lookup) and does not depend on the learning algorithm. Because admission applies to session-scoped operations, it primarily affects the frequency and latency of session-interface RPCs rather than the computational cost of local training. In practice, public keys and token parsing can be cached across requests within a session, and the gateway can be horizontally scaled as a stateless service. These properties are aligned with the deployment goal of making authorization decisions explicit at the boundary without introducing a centralized policy evaluation dependency on the critical request path.

\textit{Microbenchmark (admission overhead):} Table~\ref{tab:admission_perf} reports a microbenchmark of the admission gateway (G/W)  to provide a sanity check on the per-request cost of session admission, where ALLOW is a valid SCT+PoP, capability permitted (Positive path), DENY-Cap is a valid SCT+PoP but with capability violation (Negative case 1), and DENY-PoP is a SCT valid but with PoP binding mismatch and capability match not reached (Negative case 2).  The benchmark was run on a single host (Intel Xeon W5-3425, 24 cores, 3.20GHz base / 4.60GHz max turbo, 30MiB L3; 128GiB DDR5-4800; Ubuntu 22.04 LTS, kernel 6.8.0-94-generic HWE; Docker 27.5.1; performance governor, no concurrent load). The gateway container ran with no CPU or memory caps. Cryptographic operations use Ed25519 via libsodium 1.0.18.

\textit{Methodology:} Each iteration uses a fixed, valid SCT for a fixed \texttt{session\_id} and generates a fresh request-bound PoP (new \texttt{jti}/timestamp/nonce) over a canonical request representation. The measured path therefore covers the full runtime admission procedure: (i) token signature and claim validation; (ii) PoP signature and claim validation (\texttt{htu}, \texttt{htm}, freshness); (iii) capability matching for the requested operation; and (iv) decision-record construction. Timings are taken inside the gateway using a monotonic high-resolution clock. To avoid disk I/O and logging dominating sub-millisecond measurements, per-iteration timing records are appended to an in-memory buffer and flushed once at the end of the run; the reported numbers thus reflect admission computation rather than artifact-store persistence costs.

\begin{table}[t]
\caption{Admission G/W Performance (median over N=1000 requests)}
\label{tab:admission_perf}
\centering
\begin{tabular}{l|ccc}
\hline
\textbf{Operation} & \makecell{\textbf{ALLOW}\\\textbf{[ms]}} & \makecell{\textbf{DENY-Cap}\\\textbf{[ms]}} & \makecell{\textbf{DENY-PoP}\\\textbf{[ms]}}\\
\hline
Token verify         & \textit{0.126} & \textit{0.125} & \textit{0.124}\\
PoP verify           & \textit{0.094} & \textit{0.095} & \textit{0.124}\\
Capability match     & \textit{0.003} & \textit{0.003} & \textit{N/A}\\
Full admission check & \textit{0.261} & \textit{0.261} & \textit{0.255}\\
\hline
\end{tabular}
\end{table}

\textit{Results:} End-to-end admission is sub-millisecond and dominated by cryptographic verification (SCT signature validation and PoP verification). Capability matching is an in-memory lookup and contributes negligibly. The PoP-mismatch path short-circuits before capability evaluation (reported as N/A), but total latency remains similar because signature verification dominates. These measurements indicate that per-request admission is unlikely to materially affect the critical path of FL training and aggregation. Moreover, the reported numbers reflect single-request latency under no concurrent load, while throughput and tail latency under parallel sessions are out of scope.

\subsection{Reproducibility}
The prototype is packaged as containers and produces two classes of artifacts that support reproducibility. First, the evidence log provides request-level traces of admission decisions with reason codes and hashed bindings to canonical requests. Second, the artifact store provides run-level metadata (\texttt{run.json}) and standard FL metrics, as well as persisted model checkpoints protected by session-scoped access. Together, these artifacts enable re-checking admission decisions post hoc, correlating them with the corresponding run, and separating authorization failures from learning-related variability.

\section{Limitations and Operational Considerations}
\label{sec:limitations}

\paragraph{Horizontal FL and Session-Bound Admission} 
Our prototype and evaluation target cross-silo, \emph{horizontal FL} settings where organizations already hold local training datasets. Admission governs \emph{session-scoped operations} over these existing data holders. We do not address \emph{vertical FL} (feature-partitioned collaboration)~\cite{liu_2024} nor federated data-acquisition lifecycles covering ingestion, consent, provenance, and retention beyond session operations. Adapting the session interface and capability model to vertical FL, and extending admission beyond the session interface, are out of scope in this work.

\paragraph{Key compromise and revocation}
We assume organization-managed key custody (e.g., vault-backed) and do not treat key compromise, revocation, or recovery. Short-lived SCTs, session termination, and key rotation mitigate key-loss exposure; mid-session revocation semantics are not treated.

\paragraph{Evidence integrity model}
Decision records are stored as structured JSON with hashes to support later audit from boundary-visible artifacts (e.g., SCT, PoP). Stronger integrity mechanisms (e.g., append-only tamper-evident logs) can be substituted without changing admission semantics.

\paragraph{Evaluation scope}
Our evaluation validates the admission architecture: correct boundary enforcement, explicit failure handling, and evidence generation. It is not intended as a production-deployment study or a large-scale performance evaluation. Admission overhead is driven by per-request cryptographic verification (Table~\ref{tab:admission_perf}); practical optimizations include caching verification material and horizontally replicating the stateless admission gateway.

\paragraph{Complementary FL security mechanisms}
We do not address privacy and robustness mechanisms for training and aggregation (Section~\ref{sec:threat_model}.d).

\paragraph{Operational considerations}
Deployments must address clock skew tolerance for freshness checks and secure trust-anchor distribution. These are deployment-specific engineering constraints that we treat as operational prerequisites rather than protocol requirements.

\section{Conclusion}
\label{sec:conclusion}
Cross-silo federated learning deployments are often hindered by a practical gap in portable, and auditable authorization for session-scoped operations across organizational boundaries. We introduce a session admission layer at the session interface that uses signed SCTs and request-bound proof-of-possession PoP, enabling per-request boundary verification and structured decision records suitable for audit. Our prototype demonstrates end-to-end enforcement and evidence generation, including capability-mismatch and PoP-mismatch denial cases. The admission layer complements existing identity and IAM foundations while leaving training and aggregation procedures unchanged. Future work includes operational hardening (e.g., key rotation and revocation workflows) and broader evaluation across additional FL stacks and deployment configurations.

 
\bibliographystyle{ieeetr}
\bibliography{biblio}

\end{document}